\documentclass[12pt]{article}

\usepackage{amssymb}

\begin{document}

\begin{titlepage}

\begin{flushright}
arXiv:0905.4346
\end{flushright}
\vskip 2.5cm

\begin{center}
{\Large \bf Bounding Isotropic Lorentz Violation\\
Using Synchrotron Losses at LEP}
\end{center}

\vspace{1ex}

\begin{center}
{\large Brett Altschul\footnote{{\tt baltschu@physics.sc.edu}}}

\vspace{5mm}
{\sl Department of Physics and Astronomy} \\
{\sl University of South Carolina} \\
{\sl Columbia, SC 29208} \\

\end{center}

\vspace{2.5ex}

\medskip

\centerline {\bf Abstract}

\bigskip

Some deviations from special relativity---especially isotropic
effects---are most efficiently constrained using particles with velocities
very close to 1. While there are extremely tight bounds on some of the
relevant parameters coming from astrophysical observations, many of these
rely on our having an accurate understanding of the dynamics of these
high-energy sources. It is desirable to have reliable laboratory
constraints on these same parameters. The fastest-moving particles in a
laboratory were electrons and positrons at LEP. The energetics of the LEP
beams were extremely well understood, and measurements of the synchrotron
emission rate indicate that the isotropic Lorentz violation
coefficient $|\tilde{\kappa}_{{\rm tr}}-\frac{4}{3}c_{00}|$ must be smaller than
$5\times 10^{-15}$.

\bigskip

\end{titlepage}

\newpage

Presently, there is quite a bit of interest in the possibility that
Lorentz and CPT may not be exact symmetries of nature. If the laws of physics,
in their most fundamental form at high energies (e.g. at the Planck scale),
do not respect these symmetries, then there should be evidence (however weak) of
that fact at observable energies. There would be Lorentz- and CPT-violating
effects in the effective theory governing low-energy phenomena.

If any violation of Lorentz or CPT invariance were discovered, it would
be a breakthrough of profound importance. It would provide crucial information
about the structure of fundamental physics and clues as to what other novel effects
we could expect to uncover.
There is a parameterization of Lorentz and CPT violations in low-energy effective
field theory, known as the standard model
extension (SME). The SME contains possible Lorentz- and CPT-violating corrections
to the standard model~\cite{ref-kost1,ref-kost2} and general
relativity~\cite{ref-kost12}. Both the
renormalizability~\cite{ref-kost4,ref-colladay2} and
stability~\cite{ref-kost3} of the SME have been studied.

Historically, there have been a number of phenomenalistic or kinematic
frameworks for analyzing
the results of Lorentz and CPT tests. However, the SME has now become the standard
tool for this purpose. Since it is an effective field theory, it is useful for
parameterizing the results of a much broader variety of tests than was possible
prior to its inception.
Sensitive searches for Lorentz violation have included studies of matter-antimatter
asymmetries for trapped charged
particles~\cite{ref-bluhm1,ref-gabirelse,ref-dehmelt1} and bound state
systems~\cite{ref-bluhm3,ref-phillips},
determinations of muon properties~\cite{ref-kost8,ref-hughes}, analyses of
the behavior of spin-polarized matter~\cite{ref-kost9,ref-heckel2},
frequency standard comparisons~\cite{ref-berglund,ref-kost6,ref-bear,ref-wolf},
Michelson-Morley experiments with cryogenic
resonators~\cite{ref-muller3,ref-herrmann2}, Doppler
effect measurements~\cite{ref-saathoff,ref-lane1},
measurements of neutral meson
oscillations~\cite{ref-kost10,ref-kost7,ref-hsiung,ref-abe,
ref-link,ref-aubert}, polarization measurements on the light from cosmological
sources~\cite{ref-carroll2,ref-kost11,ref-kost21,ref-kost22},
high-energy astrophysical
tests~\cite{ref-stecker,ref-jacobson1,ref-altschul6,ref-altschul7,ref-klinkhamer2},
precision tests of gravity~\cite{ref-battat,ref-muller4}, and others.
The results of these experiments set bounds on various SME coefficients. Up-to-date
information about bounds on the SME coefficients may be found in~\cite{ref-tables};
at the present time, many of the SME coefficients are quite
strongly constrained, but many others are not.

One of the most natural ways that Lorentz violation could occur would be
to have different sectors of the standard model (e.g. photons and electrons) have
different limiting velocities
at high energies. However, an isotropic difference in the limiting
velocity for two different species turns out to be rather difficult to measure.
Direction-dependent effects can be studied by comparing the results of precision
experiments performed with the apparatus in different orientations. However, a
violation solely of
boost invariance requires a comparison of relativistic effects, which are
suppressed by two powers of the velocities involved at low energies. The greater
the velocities involved in an experiment are, the more precise are the bounds than
can be set.

The fastest-moving particles we can study are astrophysical in origin.
Astrophysical processes involving electrons and photons have been used to place
many strong constraints on SME coefficients. The observed absence of vacuum
Cerenkov radiation ($e^{-}\rightarrow e^{-}+\gamma$ with superluminal electrons),
the absence of photon decay ($\gamma\rightarrow e^{+}+e^{-}$), and the structure of
synchrotron spectra have proved particularly useful. The resulting bounds are typically proportional to $\gamma^{-2}$, where $\gamma$ is the Lorentz factor of the
massive particles involved.

However,
it is also desirable to have laboratory bounds on the SME coefficients. Many bounds
that are based on astrophysical observation rely on our having an accurate
understanding of either distant photon sources or high-energy cosmic ray air showers.
In most cases, the identities of the particles originally
responsible for producing what we actually see can only be
inferred, and these inferences may be controversial. For example, there is
disagreement whether the TeV $\gamma$-ray spectra of most sources is caused by
inverse Compton scattering or $\pi^{0}$ decay, and whether the highest energy primary
cosmic rays are protons, nuclei, or something else entirely.
If the particle identifications
are incorrect, the corresponding bounds could be completely invalidated.
However, a few astrophysical bounds do not suffer from any such such deficiency.
In particular, the conclusions drawn from the absence of the process
$\gamma\rightarrow e^{+}+e^{-}$ do not depend in any way on how the
photon involved was produced. The decay process would occur extremely rapidly, if
it were allowed. The fact that a photon reaches an Earth-based detector
without decaying provides a constraint on Lorentz violation that is completely
rigorous, in no way inferior to a measurement with photons both produced and
detected in the laboratory.

Vacuum Cerenkov radiation and photon decay are threshold phenomena. In the presence
of Lorentz violation, these ordinarily forbidden processes can occur readily. If
they are observed not to occur up to an energy $E$, the SME coefficients involved
must be smaller than ${\cal O}(m^{2}/E^{2})$, where $m$ is the electron mass.
Synchrotron radiation is more subtle;
it is ordinarily allowed, and Lorentz violation would only result in a change in
the radiation spectrum. However, this characteristic can actually be highly
advantageous. Precise monitoring of revolving particles' synchrotron losses can be
used to constrain the same SME coefficients with significantly better than
${\cal O}(m^{2}/E^{2})$ precision.

The most highly boosted particles available in a laboratory were electrons
and pos\-i\-trons at the Large Electron-Positron Collider (LEP).
Energy calibration data from LEP can be
used to place very tight constraints on isotropic Lorentz violation.
The rate of synchrotron radiation from the electrons and positrons
in the accelerator was measured with very high precision, and this fact can be
used to better constrain this extremely natural, yet poorly measured in the
laboratory, form of Lorentz violation.

The Lagrange density for the electron and photon sectors of the SME is
\begin{eqnarray}
{\cal L} & = & -\frac{1}{4}F^{\mu\nu}F_{\mu\nu}
-\frac{1}{4}k_{F}^{\mu\nu\rho\sigma}F_{\mu\nu}F_{\rho\sigma}
+\frac{1}{2}k_{AF}^{\mu}\epsilon_{\mu\nu\rho\sigma}F^{\nu\rho}A^{\sigma}
+\bar{\psi}(i\Gamma^{\mu}D_{\mu}-M)\psi \\
\Gamma^{\mu} & = & \gamma^{\mu}+c^{\nu\mu}\gamma_{\nu}-d^{\nu\mu}\gamma_{\nu}
\gamma_{5}+e^{\mu}+if^{\mu}\gamma_{5}+\frac{1}{2}g^{\lambda\nu\mu}
\sigma_{\lambda\nu} \\
M & = & m+\!\not\!a-\!\not\!b\gamma_{5}+\frac{1}{2}H^{\mu\nu}\sigma_{\mu\nu}+im_{5}
\gamma_{5}.
\end{eqnarray}
The behavior of the quanta at high energies is primarily determined by the
dimensionless, CPT-even
coefficients $c$, $d$, and $k_{F}$. They affect the velocities of
electrons, positrons, and photons. However, the $ev^{\mu}A_{\mu}$
coupling between charged particles and the electromagnetic field is not modified;
this is a consequence of electromagnetic gauge invariance, which the SME preserves.
Since the coefficients parameterizing the Lorentz violation are expected to be small,
we shall only consider their leading-order effects.

The LEP energy data is primarily sensitive to isotropic Lorentz violation, given the
bounds that have already been placed on the various SME parameters.
In the photon sector the
nineteen-component, double traceless $k_{F}$ can be separated into coefficients
that lead to photon birefringence and those which do not. The former are very
strongly constrained by cosmological
measurements~\cite{ref-kost11,ref-kost21,ref-kost22}.
Of the coefficients that are not
related to vacuum birefringence, the ones which are even under parity are already
constrained---using resonant cavity experiments~\cite{ref-muller3,ref-herrmann2}---at
roughly the $10^{-17}$ level;
the crucial exception is the isotropic
coefficient $\tilde{\kappa}_{{\rm tr}}=\frac{2}{3}(k_{F})_{\alpha}\,^{0\alpha0}$.
The parity-odd ones are less well bounded,
but they have little effect on the rate of synchrotron losses at an
accelerator. They may affect the instantaneous power emitted, but the total power
loss over a full revolution---being a parity-even quantity---is unchanged; any
increase in radiation along one side of the orbit is compensated by a decrease on
the opposite side. Therefore, of the photon-sector coefficients, only the isotropic
boost invariance violation coefficient $\tilde{\kappa}_{{\rm tr}}$ can affect the
synchrotron power at the level of interest.

In the electron sector, the $c$ coefficients affect the velocities of electrons
and positrons.
The same resonant cavity experiments used to constrain the non-birefringent $k_{F}$
coefficients can also constrain the $c$ coefficients directly, taking advantage of
the dependence of a cavity's shape on the electron-sector parameters. The parity
even $c_{jk}$ coefficients are constrained tightly enough to be neglected here, and
the parity-odd $c_{0j}$ do not affect the total power emitted during an orbit.
This leaves $c_{00}$, which is relatively poorly constrained by laboratory
experiments. However, using a coordinate redefinition,
$x'^{\mu}=x^{\mu}-c^{\mu}\,_{\nu}x^{\nu}$~\cite{ref-kost17},
it is actually possible to
eliminate $c$ from the Langragian;
only differences between the $c$ and non-birefringent $k_{F}$
coefficients are physically measurable. Although
we shall use this freedom to set $c^{\nu\mu}=0$ and henceforth only consider
$\tilde{\kappa}_{{\rm tr}}$, the bounds we shall derive will more generally
be on the combination $\tilde{\kappa}_{{\rm tr}}-\frac{4}{3}c_{00}$. The best astrophysical bounds on this quantity, disentangled from the other all other
coefficients, are $-1.3\times 10^{-14}<\tilde{\kappa}_{{\rm tr}}-\frac{4}{3}c_{00}
<8\times 10^{-15}$~\cite{ref-altschul7}. The laboratory bounds derived
here are comparable and more secure.

The $d$ coefficients are analogous to the birefringent part of $k_{F}$. Their
effects are similar to those of $c$, except that they depend on helicity and
particle-antiparticle identity. The $d$ coefficients only affect
synchrotron radiation
losses if the beams are longitudinally polarized, which can be the case
instantaneously but not over long periods. Electron helicity precesses in a
magnetic field, because of the anomalous magnetic moment. For this reason, the LEP
beam was ordinarily maintained in a transverse polarization state and only rotated
into longitudinal polarization before an interaction point.

It has recently been observed that accelerator data could be used to place new
laboratory constraints on $\tilde{\kappa}_{{\rm tr}}$~\cite{ref-hohensee1}.
The absence of vacuum Cerenkov radiation at LEP indicates that 
$\tilde{\kappa}_{{\rm tr}}<1.2\times 10^{-11}$. The fact that energetic
photons produced at the Tevatron do not decay shows that
$\tilde{\kappa}_{{\rm tr}}>-5.8\times 10^{-12}$. However, it is possible to place
much stronger bounds than these, using the same LEP energy data, by taking
advantage of the high precision to which the LEP synchrotron losses were
determined.

The synchrotron process in the presence of Lorentz violation was discussed in
detail in~\cite{ref-altschul5}. However, for the case of isotropic Lorentz
violation only, the main changes can be understood quite simply. The inclusion
of $\tilde{\kappa}_{{\rm tr}}$ in the Lagrangian changes the propagation speed of
photons to $\sqrt{\frac{1-\tilde{\kappa}_{{\rm tr}}}{1+\tilde{\kappa}_{{\rm tr}}}}
\approx 1-\tilde{\kappa}_{{\rm tr}}$. The electromagnetic sector behaves
according to ordinary special relativity, except with a modified Lorentz factor
$\tilde{\gamma}=(1-2\tilde{\kappa}_{{\rm tr}}-v^{2})^{-1/2}$. The power
radiated by a synchrotron electron is $P=\frac{e^{2}a^{2}}{6\pi m^{2}}
\tilde{\gamma}^{4}$, where $a$ is the magnitude of the acceleration; the electron
velocity is effectively increased to $v+\tilde{\kappa}_{{\rm tr}}$. (If
an electron's velocity exceeds $1-\tilde{\kappa}_{{\rm tr}}$, vacuum Cerenkov
radiation will be emitted.) For ultrarelativistic particles, $\gamma\approx
[2(1-v)]^{-1/2}$ is a very sensitive function of $v$, with $d\gamma/dv=v\gamma^{3}
\approx\gamma^{3}$. In the presence of the Lorentz violation the radiated power
becomes
\begin{equation}
\label{eq-P}
P=P_{0}(1+4\gamma^{2}\tilde{\kappa}_{{\rm tr}}),
\end{equation}
where $P_{0}$ is the radiation rate in the absence of $\tilde{\kappa}_{{\rm tr}}$.

Precise determination of the beam energy at LEP was important, since one of the
accelerator's most important functions was to provide precision measurements of
the W and Z boson masses.
The beam energy $E$ was calculated using several complementary methods.
The first method entailed measuring the magnetic field profile using nuclear
magnetic resonance (NMR) and also measuring the beam
trajectory; together these determine the beam energy. The field strength
and the radius of the orbit in the bending magnets were known to high
precision. Moreover,
the validity of the NMR measurements would not be affected by Lorentz violation; any
Lorentz violations strong enough to affect the magnetic field measurements
are ruled out by atomic clock
experiments~\cite{ref-berglund,ref-kost6,ref-bear,ref-wolf}. 

Also measured was the synchrotron tune, $Q_{s}$---the ratio of the synchrotron
oscillation frequency to the
orbital frequency~\cite{ref-assmann}. The oscillations occurred because of the
nonuniformity in the beam particles' energies. Particles with less than the
nominal beam energy revolve around smaller paths, and thus they travel between the
radio frequency (RF) accelerating
cavities more quickly. They arrive at the cavities earlier in the
RF cycle and receive larger-than-expected energy boosts. The opposite occurs for
particle with greater than the nominal energy. This effect causes synchrotron
oscillations in the beam, and a fit to their frequency provides an independent way to
determine $E$.

The fit of $Q_{s}$ produced a measurement of $E$ with a $1\sigma$ uncertainty
of 21 MeV (on a 91 GeV run)~\cite{ref-assmann}. This uncertainty was much larger
than the discrepancy of 3 MeV between the values of the energy inferred from $Q_{s}$
and from NMR. The uncertainty is primarily controlled by
the fitting uncertainty and the precision with which $Q_{s}$ is measured.
$E$ and $U_{0}$ (the energy loss per revolution) enter the formula for
$Q_{s}$ through $(g^{2}e^{2}V_{RF}^{2}-U_{0}^{2})/E^{2}$, and under the steady
state conditions at which the collider operated, $U_{0}=
geV_{RF}\sin\psi_{s}$. $V_{RF}$ and $\psi_{s}$ are the amplitude and phase of the RF
voltage during the
beam's passage through the accelerating cavities; both are known to high precision.
$g$ is a correction factor related to possible phasing errors and misalignments of
the cavities. It is a fit parameters and a significant source of
uncertainty; however, in absolute terms, its value is close to 1, and its value can
be determined using
a separate fit to $Q_{s}$ performed well below real experimental energies.
The presence of $\tilde{\kappa}_{{\rm tr}}$ does not change the formula for $Q_{s}$,
except through a
rescaling of $U_{0}$; since the Lorentz force law is not modified, the
motion of the particles in the applied fields and the energy imparted by the RF
cavities are unchanged.

$U_{0}$ was not a parameter that was varied in the
fit; it was assumed to take the conventional synchrotron form, corrected to account
for additional well understood
losses (primarily related to finite beam size and parasitic
impedance interactions; these and other corrections were either modeled from first
principles or measured directly).
However, since $E$ was independently and more accurately known from NMR
measurements, it is possible to reinterpret the fit for $E$ as a fit for $U_{0}$.
Since $E$ and $U_{0}$ enter the formula for $Q_{s}$ only in the combination
$-(U_{0}/E)^{2}\cos^{2}\psi_{s}$, the uncertainty ascribed to
$E$ in the fit for the energy is equivalent to essentially the same
fractional uncertainty of $2.4\times 10^{-4}$ in $U_{0}$ (and hence $P$).

Conservatively, we may state that the fractional deviation of $P$ from its
conventionally
expected value is $\eta <6\times 10^{-4}$, for measurements performed at the Z
pole energy of 91 GeV (corresponding to $\gamma>1.7\times10^{5}$). This represents
a $2\sigma$ bound, and it accounts
for all additional sources of error, such as the error in the NMR measurement of $E$
and the discrepancy between that measured value and the value inferred from
$Q_{s}$.
Then according to
(\ref{eq-P}),
\begin{equation}
|\tilde{\kappa}_{{\rm tr}}|<\frac{\eta}{4\gamma^{2}}<5\times10^{-15}.
\end{equation}

This bound is not as strong as the bound on $\tilde{\kappa}_{{\rm tr}}$ that
comes from the absence of photon decay~\cite{ref-stecker}. The absence of
$\gamma\rightarrow e^{+}+e^{-}$ for up to 50 TeV photons gives a bound at the
$2\times 10^{-16}$ level, although that bound is strictly one-sided; only negative
values of $\tilde{\kappa}_{{\rm tr}}$ are so constrained. Moreover, the photon
decay bounds are also entangled with bounds on the parity-odd coefficients.
Therefore, the current result represents an improvement in clean, reliable,
laboratory-derived bounds of three orders of
magnitude. This improvement over the vacuum Cerenkov bounds comes precisely from
the $\lesssim10^{-3}$ precision with which the synchrotron loss rate is known.

While the parity-odd coefficients $\tilde{\kappa}_{o+}$ (or equivalently $c_{0j}$)
do not contribute to the energy lost during a full revolution, they do affect the
instantaneous rate of of synchrotron emission. There is an additional,
direction-dependent modification of the speed of light, which, in the presence of
a generic non-birefringent $k_{F}$ is
$1-\frac{1}{2}\left[\tilde{k}_{jk}\hat{v}_{j}\hat{v}_{k}+
2\tilde{k}_{0j}\hat{v}_{j}+\tilde{k}_{00}\right]$,
where $\tilde{k}^{\mu\nu}=(k_{F})_{\alpha}\,^{\mu\alpha\nu}$. The
$\tilde{k}_{0j}$ are equivalent to $\epsilon_{jkl}(\tilde{\kappa}_{o+})^{kl}$
and the $\tilde{k}_{jk}$ to $(\tilde{\kappa}_{e-})_{jk}$. These generalize the
isotropic case, which has $\tilde{k}_{00}=\frac{3}{2}
\tilde{\kappa}_{{\rm tr}}$ and $\tilde{k}_{jk}=\frac{1}{2}\tilde{\kappa}_{{\rm tr}}
\delta_{jk}$. With more detailed data on the
emission rates of the LEP electrons and positrons as they moved, one could
potentially place
bounds (comparable to the $\tilde{\kappa}_{{\rm tr}}$ bounds) on the
$\tilde{\kappa}_{o+}$ parameters.

In summary, we have derived new bounds, based on terrestrial laboratory
experiments, on the isotropic Lorentz violation parameter
$\tilde{\kappa}_{{\rm tr}}$. The precision with which synchrotron losses at LEP
matched conventional expectations constrains electron and photon SME parameters
to be $|\tilde{\kappa}_{{\rm tr}}-\frac{4}{3}c_{00}|<5\times 10^{-15}$.
This new two-sided
constraint represents an improvement of three orders of magnitude over previous
laboratory bounds on these quantities. It also represents a modest improvement over
the best astrophysical bounds on $\tilde{\kappa}_{{\rm tr}}-\frac{4}{3}c_{00}$
alone.

\section*{Acknowledgments}
The author is grateful to M. Purohit for helpful discussion.

\end{document}